# Towards Evaluating the Quality of a Spreadsheet: The Case of the Analytical Spreadsheet Model

Thomas A. Grossman

Vijay Mehrotra

Johncharles Sander

tagrossman@usfca.edu, vmehrotra@usfca.edu, jgsander@dons.usfca.edu

University of San Francisco, School of Management

2130 Fulton Street, San Francisco, CA  94117-1045, USA

**ABSTRACT**

We consider the challenge of creating guidelines to evaluate the quality of a spreadsheet model. We suggest four principles. First, state the domain—the spreadsheets to which the guidelines apply. Second, distinguish between the process by which a spreadsheet is constructed from the resulting spreadsheet artifact. Third, guidelines should be written in terms of the artifact, independent of the process. Fourth, the meaning of "quality" must be defined. We illustrate these principles with an example. We define the domain of "analytical spreadsheet models", which are used in business, finance, engineering, and science. We propose for discussion a framework and terminology for evaluating the quality of analytical spreadsheet models. This framework categorizes and generalizes the findings of previous work on the more narrow domain of financial spreadsheet models. We suggest that the ultimate goal is a set of guidelines for an evaluator, and a checklist for a developer.

# 1. INTRODUCTION

This paper seeks to advance our understanding of a very important question: If I open a spreadsheet I have not seen before, how can I determine whether it is a "good spreadsheet"? From the literature and conversations with practitioners, there is no clear agreement regarding this question.

Discussions at EuSpRIG conferences over the years have approached this question in many ways, sometimes at a high level and sometimes at a very detailed level, but never at a level where clear and coherent agreement could be achieved. It is evident that the apparently simply question of the quality of a spreadsheet is, in fact, very complex. Indeed, the only agreed-upon answer to "is this a good spreadsheet?" would seem be the two words "It Depends"!

The purpose of this paper is to change the conversation regarding "what makes a good spreadsheet". We think that spreadsheet experts agree more than they disagree. In this paper, we make an attempt to identify the areas of agreement and construct a framework that, we hope, can encompass existing spreadsheet development methodologies and recommendations. Our ultimate goal is to unify the many perspectives, approaches, assumptions, and special cases that exist in the literature and in the hard-won habits of practitioners, to provide a springboard for detailed methodologies for specific types of spreadsheets.

## 1.1. Overview

We start by articulating a set of principles that we believe are essential for evaluating the quality of a spreadsheet. We then apply the principles to a specific situation, the domain of analytical spreadsheet models. In the conclusions we discuss our hopes for a dialog around the ideas in this paper, and where the work might go next.

In section 2 we propose four principles related to evaluating spreadsheet quality:

1. Any guidelines require a clear definition of where they apply; that is, one must define the "domain" of spreadsheets to which the guidelines apply.

2. The meaning of "quality" must be defined.

3. It is essential to distinguish between the spreadsheet "artifact" and the "process" by which that artifact was created.

4. Spreadsheet quality should be evaluated in terms of the spreadsheet artifact, not the process.

We believe that these principles are essential for any guidelines for evaluating spreadsheet quality.

In sections 3 – 12 we apply the principles for a specific class of spreadsheet. In section 3 we define the domain of "analytical spreadsheet model" and explain the quality criteria that we use. In section 4 we discuss the wisdom of practice that informs the following sections. In section 5we provide an overview of an initial framework for evaluating the quality of an analytical spreadsheet model. In sections 6 – 12 we discuss elements of the framework.

In the final section, we discuss where we might go next. In particular, it is our hope to develop useful, detailed guidelines for analytical spreadsheet models that are broadly consistent with the existing guidelines for narrow sub-domains such as large-scale financial planning models.

**1.2. Contribution**

This paper makes the following contributions.

1. Proposes four principles for evaluating the quality of a spreadsheet.

2. Defines a new domain of analytical spreadsheet models

3. Provide a rigorous framework and terminology for discussing the design quality of an analytical spreadsheet model.

4. Shows how the existing more narrow sub-domain of large-scale financial planning models fits into this domain.

5. Generalizes the existing literature and practice wisdom for large-scale financial planning models to the broader domain of analytical spreadsheet models

6. Hypothesizes on the nature of future contributions, specifically guidelines for evaluating the quality of analytical spreadsheet models, and a checklist for the developers of such models.

This paper does not propose a detailed evaluation methodology, as it would be premature to do so. We must first get the framework right. This paper does not provide an example of evaluating the quality of a specific spreadsheet because that would be premature.

**2. PRINCIPLES FOR EVALUATING SPREADSHEET QUALITY**

We propose four principles for any evaluation of spreadsheet quality.

**2.1. A Clear Domain Definition is Required**

The world of spreadsheets is vast and diverse (for example, see Jelen 2005, Grossman, Mehrotra and Özlük 2007, and Powell, Baker and Lawson 2009), ranging from business models to quilting patterns to decorative art. It is difficult to imagine a set of universal guidelines that apply to all spreadsheets. The specificity and usefulness of guidelines increases as the scope of spreadsheets under consideration decreases. Thus, any recommendations will be useful only for a limited set (or "domain") of spreadsheets.

Therefore, it is essential that a clear domain definition be provided for any spreadsheet guidelines. One of the characteristics of the human experience is a tendency for the word "spreadsheet" to be used to mean "the types of spreadsheets that I and my friends tend to build". This is perfectly natural, but for the field of spreadsheet engineering to advance it is necessary that guidelines and methodologies carefully define the domain to which they apply. However, the existing literature is characterized by a lack of clean domain definitions and this confounds our ability to understand where methodologies apply.

**2.2. Define What is Meant by "Quality"**

It is well known in the discipline of operations management that the term "quality" has many different meanings. For example, "quality" can refer to conformance to specification, compliance with tolerances, fitness for use, a count of defects, the extensiveness of features, cost, exclusivity of access, or other meanings. Thus, any activity related to quality must carefully define what aspect of quality is of interest.

This definition of quality will depend upon the purpose of the spreadsheet and the goals of the evaluator. For example, a quilting pattern spreadsheet might be evaluated on how easily it can be observed while sewing, whereas a financial model might be evaluated on accuracy and how easily it can be modified. The relevant quality dimensions will depend upon the purpose of the spreadsheet and the goals of the evaluator. It is essential that these quality dimensions not be assumed obvious or otherwise left unstated. Any evaluation of spreadsheet quality be preceded by an explicit definition of what dimensions of quality are to be evaluated.

### 2.3. Distinguish between the Artifact and the Process

The most powerful spreadsheetdevelopment methodologies are those promulgated for large-scale financial planning models (FAST 2010, Swan 2008, and SSRB 2005, compared in Grossman and Özlük 2010). These very detailed methodologies provide process guidelines for the development of spreadsheet models. The power of these methodologies arises in part because the spreadsheet itself is intimately connected with the way that it was constructed; the layout is intended to support construction, and the construction is intended to build the layout. We hypothesize that a rigorous, detailed process is only possible for a rigorous, standardized design.

However, this tight coupling between process and the artifact that results from the process makes it difficult to compare methodologies. The quality of a spreadsheet model – the artifact – is viewed through the lens of the design process. The effectiveness of the design process is viewed through the lens of its ability to rapidly and robustly create the standardized design. In order to achieve any level of generality in guidelines, it is essential that we separate our thinking about the process of creating a spreadsheet and our studyof the spreadsheet artifact. We must distinguish between the "what" and the "how".

### 2.4. Evaluate Quality in Terms of the Spreadsheet Artifact

Any useful discussion of spreadsheet quality must be flexible enough to allow for different designs. This means that we must break away from the constraints of process. Spreadsheet quality needs to be defined as an attribute of the spreadsheet, without any knowledge of how the spreadsheet was constructed.

For devotees of a deep spreadsheet development methodology, this can be a challenge, because the quality of a spreadsheet model – the artifact – is naturally considered in the context of the construction methodology. This serves to confuse the evaluation of spreadsheet quality, because quality is in part defined as adherence to an *a priori* standardized design. We need to define quality independent of the constraints of any particular design.

## 3. THE CASE OF THE ANALYTICAL SPREADSHEET MODEL

We illustrate the above principles for a particular domain (principle 1) and for a particular set of quality criteria (principle 4). We limit our examination to the spreadsheet artifact, and do not include any process guidelines (principles 2 and 3).

### 3.1. Domain Definition: The Analytical Spreadsheet Model

We start by defining the domain. We then present a framework that we intend to be helpful for evaluating the quality of any analytical spreadsheet model, regardless of how it was constructed.

The domain is based on our experience teaching in American business schools where we see spreadsheet models grounded in the discipline of management science, as well as models in finance, marketing, supply chain, and other business disciplines. These models, to us, have much in common and we find that our advice to students is similar across a variety of models. Therefore we define the domain as follows.

We define an *analytical spreadsheet model* as(1) a spreadsheet computer program that (2) implements a mathematical model (3) for purposes of analysis that (4) serves as an organizational asset that is (5) employed in a larger business context. An analytical spreadsheet model has the following properties:

1. As a *spreadsheet computer program*, the analytical spreadsheet model is programmed in a spreadsheet language such as Excel.

2. An analytical spreadsheet model implements a conceptual *mathematical model* [Powell and Baker 2010, Grossman and Özlük 2004] that articulates domain expertise.

3. The analytical spreadsheet model exists to enable *analysis* upon it by changing inputs and observing outputs. [Grossman 2008, Powell and Baker 2010, Spreadsheet Analytics 2010].

4. As an *organizational asset*, the analytical spreadsheet model is intended for use not only by the author but also by the author's immediate colleagues and/or successors; usage could be intensive use for a short period, or routine use over time.

5. The analysis is performed in the context of a *larger business context* that has important organizational goals, hence the outputs of the spreadsheet are designed to meet the needs of people besides the author.

We note that an analytical spreadsheet model is different than a "data-driven" spreadsheet model that starts with a large set of numeric values, and then seeks to extract insight from them. A spreadsheet that relies heavily on business intelligence tools such as sort, filter, and pivot table will not normally be considered an analytical spreadsheet model.

An analytical spreadsheet model can be constructed purposefully, or it can arise without intention as a "legacy spreadsheet application" (Grossman, Mehrotra and Özlük 2007).The level of spreadsheet engineering investment in an analytical spreadsheet model is more than that required for an informal "quick and dirty model" or "personal productivity tool", but less than that required for a formal "spreadsheet application" [Grossman 2007] that is to be deployed to multiple less-sophisticated users.

Note that analytical spreadsheet models are primarily about the business logic embedded in the model. Models that are primarily data-driven, with a large dataset and relatively few cell formulas do not fall in this domain.

Analytical spreadsheet models are used in business, including supply chain management, market research, worker scheduling, finance, as well as in engineering and science. Analytical spreadsheet models are taught extensively in business school courses on management science and quantitative analysis. Examples of analytical spreadsheet models across many areas of business can be found in Powell and Baker 2010, Winston & Albright 2008, Ragsdale 2010, and Interfaces 2008, and a rich variety of finance examples can be found in Benninga 2008. The spreadsheets described in Read and Batson 1999 and Tennent and Friend 2001, and the large-scale financial planning models of FAST 2010, Swan 2008, and SSRB 2005 are analytical spreadsheet models. Most or all of the spreadsheets discussed in Croll 2005 are analytical spreadsheet models. Grossman,

Mehrotra and Özlük 2007 describe analytical spreadsheet models across diverse industries. We indicate the domain of analytical spreadsheet models in Figure 1.

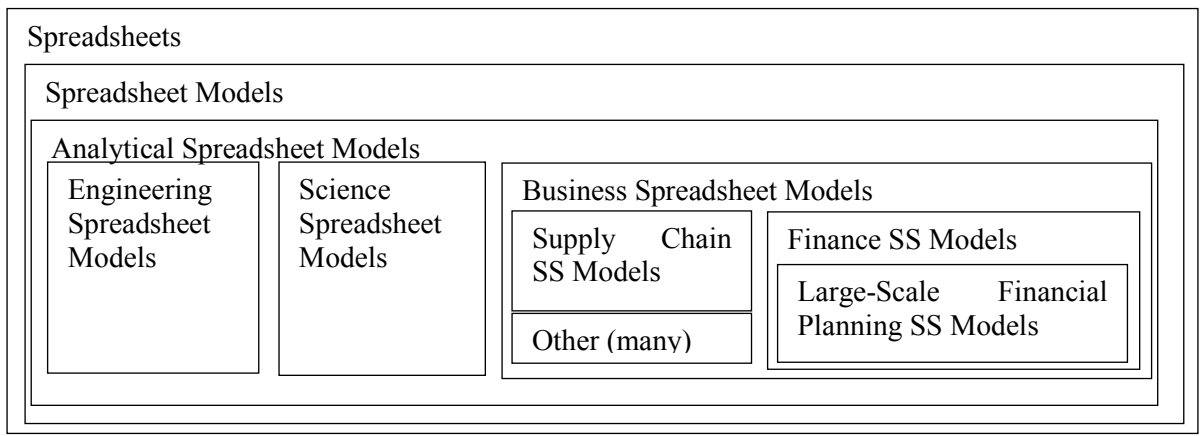

**Figure 1: Venn Diagram for Analytical Spreadsheet Models**

We emphasize that this paper considers the spreadsheet as an artifact; we consider the quality of the spreadsheet model in isolation from the manner in which it was created. Thus, we do not address the *process* by which the spreadsheet is constructed. For example, we do not discuss recommendations for how to go about creating cell formulas, nor how to test or code-inspect a spreadsheet to detect any errors. We note that our guidelines are consonant with the process recommendations of FAST 2010, Swan 2008, and SSRB 2005.

**3.2. Criteria for Quality**

We are interested in the properties of a "high-quality" analytical spreadsheet model. What are the "right" quality dimensions for an analytical spreadsheet model? This question will undoubtedly spark debate. In the interests of initiating a constructive discussion, we propose that the following quality dimensions. A high-quality analytical spreadsheet model must be:

1. Suitable for efficient analysis
2. Readable (can be understood by a non-author; see McConnell p. 842)
3. Transferable (to another analyst or the author's successor)
4. Accurate (compute the mathematical model without error)
5. Reusable (can easily be employed again with different data)
6. Modifiable (can be extended or adapted to new circumstances)

These six quality dimensions come directly from the definition of the analytical spreadsheet model. They are, in our view, necessary (we repeat the definition of an analytical spreadsheet model) for a spreadsheet computer program that implements a mathematical model for purposes of analysis that serves as an organizational asset that is employed in a larger business context.

We believe that obtaining agreement on the dimensions of quality will aid in the creation of a guidelines and recommendations. In the framework of guidelines that appears in later sections, each element is devised to contribute to one or more of these quality dimensions.

## 4. LITERATURE AND THE WISDOM OF PRACTICE

This paper started when we sought to articulate for the benefit of our business students what makes a high-quality analytical spreadsheet. There is no empirical research on this topic. Therefore, as in many practical disciplines, these guidelines are based in the scholarly tradition of "the wisdom of practice"(see Weimer 2006). This paper articulates the wisdom of practice from the literature and our experience, along with our tacit knowledge accumulated from years of experience in industry and academia engaging in business consulting, modeling, software development, and spreadsheet engineering.

We integrate the published results from the spreadsheet engineering community (especially the European Spreadsheet Risks Interest Group), and draw upon the principles of the mainstream software engineering literature. It is our intention that the framework presented here is consistent with the recommendations for business models in Read and Batson 1999 and financial models in Tennent and Friend 2001, and the detailed guidelines for large-scale financial planning models of FAST 2010,Swan 2008, and SSRB 2005, summarized in Grossman 2010.

## 5. FRAMEWORK FOR EVALUATING THE QUALITY OF AN ANALYTICAL SPREADSHEET MODEL

Our ultimate goal is a set of practical, useful guidelines for evaluating the quality of an analytical spreadsheetmodel. These guidelines should be specific enough to be useful, but broad enough to encompass the detailed recommendations of the existing methodologies. We believe that any set of useful guidelines for evaluating the quality of an analytical Spreadsheet model should address the following elements:

- Modular design
- Structured design
- Design for input-output usage
- Disciplined information flow
- Distinct inputs module(s) with certain properties
- Distinct model computations module(s) with certain properties
- Distinct reports module(s) with certain properties

We provide a sketch of each of these elements in the following sections.

## 6. MODULAR DESIGN

A *module* is a set of similar things. A module can contain subsidiary modules (sub-modules). A module can be a few cells; a portion of a worksheet; an entire worksheet; a set of worksheets; or an entire workbook. A top-level module is a module that is not a sub-module. A bottom-level module is a module that contains no sub-modules.

Each module should have a clear purpose. A module with distinct activities contains sub-modules, each with a clear purpose.

The need for input-output usage discussed above requires that there be three distinct top-level modules: Inputs, Model Computations, and Reports. The Inputs module receives all model inputs in raw form, and adapts them as necessary to the needs of the Model Computations module. The Model Computations module performs the computations necessary to implement a mathematical model that is known to the author. The Reports module presents model results in a form that is convenient to the consumer.

## 7. STRUCTURED DESIGN

*Structured design* is the purposeful arrangement of the analytical spreadsheet model into a set of connected building blocks called "modules". Structured design is a standard technique from traditional software engineering used to manage complexity. Structured design is essential for readability, accuracy, reusability, and modifiability.

In addition, structured design for an analytical spreadsheet model supports input-output usage; is modular; has disciplined information flow; and comprises an Inputs module, a Model Computations module, and a Reports module.

## 8. DESIGN FOR INPUT-OUTPUT USAGE

For efficient analysis using an analytical spreadsheet model, the analyst must be able to focus on the values for model inputs and resulting model outputs, without being distracted by "internal" model calculations. (I.e., the analyst interacts with the analytical spreadsheet model the same way he interacts with traditional shrink-wrap software, a decision support system, or web analytics software.) The spreadsheet must be designed so that:

- analysis does not involve observation of the model logic.
- inputs are grouped together separate from calculations.
- outputs are grouped together separate from calculations.

Design for input-output usage supports suitability for analysis, and transferability.

Note that design for input-output usage implies that a "calculator" design, where inputs are entered then used for calculation, and more inputs are entered then used for calculation, is unsuitable for efficient analysis. Such a spreadsheet should be converted to an input-output design prior to service as an analytical spreadsheet model.

## 9. DISCIPLINED INFORMATION FLOW

Information flows through the modules via a direct, non-circuitous path. At the top level, information flows from the Inputs module to the Model Computations module to the Reports module.

Disciplined information flow is summarized in the structure chart [Structure Chart 2011] of Figure 2.

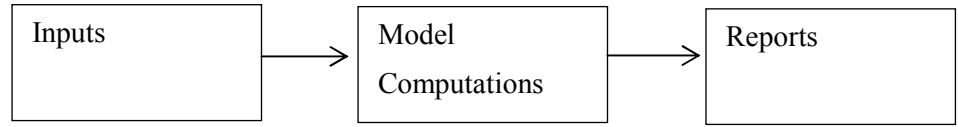

Figure 2: Structure Chart Showing Information Flow Among Top-Level Modules

Disciplined information flow insures suitability for input-output usage, and supports modifiability as any future changes to the model logic in the Model Computations module are isolated from the Inputs and Reports.

## 10. INPUTS MODULE

The Inputs module manages all the inputs necessary for model computations. The design and construction of the Inputs module is a spreadsheet engineering exercise that requires little domain expertise, although some domain expertise can be required for quality assurance.

The Inputs module receives all model inputs in raw form, and adapts them as necessary to the needs of the Model Computations module. The design of the Inputs module can have sub-modules.

We would suggest that it is useful to distinguish among sub-modules. Based on experience with a wide variety of models, we suggest that is valuable to provide as necessary distinct sub-modules for different types of inputs including *Source Data*(external inputs that are outside the organization's control, presented in their original form); *Assumptions* (inputs that are judgmental, tendentious, or frequently change) ; *Decision Variables* (business choices under the organization's control);and *Input Pre-Processing* (changes to the other inputs to make them more useful). A spreadsheet model may include other modules as appropriate.

## 11. MODEL COMPUTATIONS MODULE

The Model Computations module performs the computations necessary to implement a mathematical model that is known to the spreadsheet developer. The logic of the model resides in the Model Computations module. The Model Computations module is the place where the domain expertise of the developer is expressed as a spreadsheet computer program. Tennent and Friend 2001 and Swan 2008 refer to the Model Computations module as the "workings".

The Model Computations module takes information from the Inputs module, performs computations, and ultimately generates the information required by the Reports module.

What might be flexible, robust guidelines for the Model Computations module? It should allow for one or more worksheets; this could be generalized as sub-modules for different blocks of computations, which would allow multiple worksheets or only a single worksheet. There should undoubtedly be a guideline related to the arrangement of inputs to make them most suitable for computation, or perhaps bringing ("echoing" or "linking") distant cells to a place close to the computation, to generalize the concept of "ingredients" from FAST 2010.

## 12. REPORTS MODULE

The Reports module contains all the outputs from the model. The Reports module and sub-modules should be designed and formatted to meet the needs, preferences, and habits of the consumer. Its design may require little or no domain expertise, although domain expertise can be required when devising new, non-standard reports.

The Reports module should do no or only trivial computations, and contain primarily links to cells in the Inputs and Computations modules.

## 13. CONCLUDING COMMENTS

Our intention is to advance the way we think about the quality of spreadsheets to recognize that diversity in spreadsheets requires multiple, focused evaluation schemes; to make more general certain existing, narrow spreadsheet engineering recommendations, and ultimately provide useful guidelines for evaluating quality. Our intent is to provide the basis for constructive discussion that will lead, in time, to a framework and set of actionable guidelines for analytic spreadsheet models, and also provide an approach for devising guidelines for other domains of spreadsheet models.

We believe that when evaluating spreadsheet quality it is essential to observe four principles. 1) Define the domain. 2) Define what is meant by "quality" prior to evaluating it. 3) Distinguish between the spreadsheet-as-artifact and the process of constructing the spreadsheet. 4) Evaluate spreadsheet quality based only on the spreadsheet itself. We wonder whether there might be other such principles.

We defined a domain of "analytical spreadsheet models". We indicated the quality dimensions that seem to be important. We ask whether these are in some sense the "right" quality dimensions for this class of models. We then provided a framework for evaluating the quality of an analytical spreadsheet model. This framework obviously requires more detail, but before we add we must first ask whether the elements of the framework are correct.

The next steps are to finalize the framework, add detail, and consider how each element can be meaningfully evaluated or measured. In addition, further work will be required to consider what spreadsheet engineering properties should be in place to ensure reusability, for example cell protection and other coding details which might or might not be considered appropriate, depending on circumstances that need to be circumscribed.

It is our hope that the newly-defined domain of analytical spreadsheet models and the framework serve as a "broad church" that encompasses best practices from the literature. Where it is perceived not to do so will represent powerful opportunities to advance our knowledge.

We emphasize that we consider in this paper only the evaluation of the artifact—the quality of an analytical spreadsheet model. Further research is needed to provide general guidelines on other aspects of the spreadsheet engineering of analytical spreadsheet models, including construction approaches and the mechanics of entering cell formulas and other information into the spreadsheet, and design aspects such as style, visual formats, descriptors, labels, units of measure, titles, authorship information, cell formulas, choice of functions, source code integrity, and visibility.